\documentclass[12pt,3p]{elsarticle}

\usepackage{graphicx}
\usepackage[T1]{fontenc}
\usepackage{amsmath}
\usepackage{mathtools}
\usepackage{amsthm}

\usepackage{amssymb}

\makeatletter
\def\ps@pprintTitle{%
 \let\@oddhead\@empty
 \let\@evenhead\@empty
 \def\@oddfoot{}%
 \let\@evenfoot\@oddfoot}
\makeatother

\begin{document}

\begin{frontmatter}

\title{Ornstein-Uhlenbeck Process with Fluctuating Damping
}

\author[label1]{Chai Hok Eab}
\address[label1]{386/23 Phetchbury 14,Bangkok 10400, Thailand}
\ead{Chaihok.E@gmail.com}

\author[label5]{S.C. Lim\corref{cor1}}
\address[label5]{Faculty of Engineering, Multimedia University
 63100 Cyberjaya, Selangor Darul Ehsan, Malaysia}
\ead{sclim47@gmail.com}
\cortext[cor1]{corresponding author}

\begin{abstract}
This paper studies Langevin equation with random damping due to multiplicative noise and its solution. 
Two types of multiplicative noise, namely the dichotomous noise and fractional Gaussian noise are considered. 
Their solutions are obtained explicitly, with the expressions of the mean and covariance determined explicitly.
Properties of the mean and covariance of the Ornstein-Uhlenbeck process with random damping, 
in particular the asymptotic behavior, are studied. 
The effect of the multiplicative noise on the stability property of the resulting processes is investigated.
\end{abstract}


\end{frontmatter}



\section{Introduction}
\label{sec:introduction}
Studies of harmonic oscillator with randomly varying parameters dated back several decades ago. 
Disordered linear oscillator chain with random frequency or random mass was first considered by Dyson in 1953 
\cite{Dyson53}.
Oscillator with random parameters has been studied in subsequent work 
\cite{Leibowitz63,Frisch68,BourretFrischPouquet73,KlyatskinTatarskii74,VanKampen76}.
Since then, a considerable amount of work covering oscillator with random frequency, 
random damping and random mass has appeared. 
A comprehensive review of the past work can be found in the book by 
Gitterman \cite[and references therein]{Gitterman14}.

Motivations for studying oscillators with fluctuating parameters come from 
their potential applications in modeling many natural and man-made systems. 
Brownian motion of a harmonic oscillator with random mass has been used to model systems 
where particles of the surrounding medium not only collide with Brownian particles, 
they can also adhere to them. 
Examples of such applications include diffusion of clusters with randomly growing masses 
\cite{Luczka95,AusloosLambiotte2006a},
planet formation by dust aggregation 
\cite{BlumWurmKempfEtAl2000,BlumWurm00,WeidenschillingEtAl97},
cluster dynamics during nucleation 
\cite{KuipersBarkema2009},
[13], 
growth of thin film 
\cite{Kaiser02},
deposition of colloidal particles on an electrode 
\cite{PerezSavilleSoria01},
traffic flow 
\cite{Nagatani96,Ben-NaimKrapivskyRedner1994}, 
stock market prices 
\mbox{\cite{Ausloos2002,Canessa2009}},
etc. 

An oscillator in addition to the possibility of having random mass, 
it can also acquire fluctuating frequency. 
The influence of the fluctuating environment such as the presence of colored noise and viscosity can be reflected 
in fluctuating damping term or random oscillator frequency. 
Examples of applications of such processes include propagation and scattering of waves in a random medium 
\cite{KlyatskinTatarskii74,Klyatskin2005}
and turbulent ocean waves, 
low amplitude wind-driven waves on the ocean surface 
\cite{Tatarski61,Phillips1977,WestSeshadri1981},
financial markets in econophysics 
\cite{TakayasuSatoTakayasu1997}
and population dynamics in biology 
\cite{Turelli1977,%
VitanovarXiv2013,VitanovVitanov2014}.
Recently, oscillator with fluctuating frequency has been used in the study of nano-mechanical resonators, 
where frequency fluctuations are resulted from molecules randomly adsorbing and desorbing onto or from the resonator, 
or diffusing along its surface 
\cite{DykmanKhasinPortmanEtAl2010,ZhangMoserGuettingerEtAl2014,MiaoYeomWangEtAl2014,%
SnyderJoshiSerna2014,SunZouMaizelisEtAl2015,SansaSageBullardEtAl2016}.

The equations that describe one-dimensional Brownian motion of a free particle in a fluid are given by
\begin{subequations}
\label{eq:introduction_0010}
\begin{align}
  \frac{dx(t)}{dt}  & = v(t), \label{eq:introduction_0010a}\\
  m \frac{dv(t)}{dt} + \gamma v(t)  & = \eta(t), \label{eq:introduction_0010b}
\end{align}  
\end{subequations}  
Here $m$ is the particle mass, and
$\gamma$ is the dissipative parameter of the viscous force of the fluid, and
$\eta(t)$ is the random force due to the density fluctuation of the surrounding medium.
(\ref{eq:introduction_0010a}) and (\ref{eq:introduction_0010b})
are known as the Langevin equations for the Brownian particle. 
The random force for the usual Langevin equation is given by the Gaussian white noise with zero mean and covariance
\begin{align}
  \bigl<\eta(t)\eta(s)\bigr> & = \delta(t-s).
\label{eq:introduction_0012}
\end{align}  
Note that (\ref{eq:introduction_0010b}) is also known as the Langevin equation for the Ornstein-Uhlenbeck velocity process $v(t)$.

For the Brownian motion of a damped harmonic oscillator driven by white noise, one has
\begin{align}
   m\frac{d^2x(t)}{dt^2} + \gamma \frac{dx(t)}{dt} + \omega^2 x(t) & = \eta(t),
\label{eq:introduction_0011}
\end{align}
where $\omega$ is the intrinsic oscillator frequency.
If the viscous damping force 
$-\gamma v(t)$
is much larger than the inertial term 
$md^2x/dt^2$ in (\ref{eq:introduction_0011}) 
such that to a good approximation the first term in (\ref{eq:introduction_0011}) can be neglected.
Such an overdamped limit results in a stochastic differential equation for the position $x(t)$:
\begin{align}
  \frac{dx(t)}{dt} & = -\mu x(t) + \zeta(t),
\label{eq:introduction_0040}
\end{align}
where $\mu = \omega^2/\gamma$, and $\zeta(t) = \eta(t)/\gamma$.
Equation \mbox{(\ref{eq:introduction_0040})}, which has the same form (except for the constant parameters) as the Langevin equation
for the velocity of free Brownian motion (\ref{eq:introduction_0010b}), can be regarded as the Langevin equation of a diffusion
process in a harmonic oscillator potential
$V(x) = \mu x^2/2$.
The solution of 
\mbox{(\ref{eq:introduction_0040})} known as the Ornstein-Uhlenbeck position process is stationary in the large-time limit. 
This stationary Ornstein-Uhlenbeck process can be associated with the quantum mechanical oscillator. 
It is mainly for this reason that the stationary Ornstein-Uhlenbeck process is also known as oscillator process 
\cite{Simon79,Roepstorff94,Dimock11,Casteren15},
especially in the books on path integral formulation of quantum theory 
\cite{Simon79,Roepstorff94,Dimock11}.

The constant coefficient stochastic differential equation (\ref{eq:introduction_0011}) for 
a damped harmonic oscillator driven by white noise can be generalized to 
\begin{align}
    m(t)\frac{d^2x(t)}{dt^2} + \gamma(t) \frac{dx(t)}{dt} + \omega^2(t) x(t) & = \eta(t), 
\label{eq:introduction_0020}
\end{align}
where $m(t)$, $\gamma(t)$ and $\omega(t)$ can be deterministic or random functions of time. 
The case  $m(t)$, $\gamma(t)$ and $\omega(t)$ are deterministic functions has been studied by many authors 
(see for examples 
\cite[and references therein]{DantasPedrosaBaseia1992,DantasPedrosaBaseia1992a,%
KanasugiOkada1995,ChaJangJungEtAl2015%
}).
On the other hand, it is possible that the mass, 
damping coefficient and frequency can be fluctuating functions of time. 
The randomness may come from the random mass, fractal structure of the medium or random orientation of the Brownian particle, 
the external effect due to viscosity. 
The external randomness may cause by thermodynamic, electromagnetic and mechanical sources. 
This exterior fluctuation is crucial for random frequency. 
Gitterman \cite{Gitterman14} has provided a comprehensive discussion on oscillator with random mass, random damping and random frequency.

Harmonic oscillator with random damping and random frequency lately has attracted renewed interest 
due partly to its potential application to the modeling of certain nano-devices such as nanomechanical resonators and nanocantilever 
\cite{ZhangMoserGuettingerEtAl2014,MiaoYeomWangEtAl2014,%
SnyderJoshiSerna2014,SunZouMaizelisEtAl2015,%
SansaSageBullardEtAl2016%
}.
The case of harmonic oscillator with random damping 
\cite{TakayasuSatoTakayasu1997,JasnowGerjuoy1975,%
GrueOeksendal1997,LuczkaTalknerHaenggi2000,%
Gitterman2004,Gitterman2005,
LeprovostAumaitreMallick2006,MendezHorsthemkeMestresEtAl2011%
}
and random frequency 
\cite{BourretFrischPouquet73,WestLindenbergSeshadri1980,%
Lindenberg1980,LindenbergSeshadriWest1980,%
LindenbergSeshadriWest1981,Gitterman2003,MankinLaasLaasEtAl2008%
}
given by white noise and dichotomous Markov noise (the random telegraph process) is well-studied 
\cite{BourretFrischPouquet73,JasnowGerjuoy1975,%
GrueOeksendal1997,LuczkaTalknerHaenggi2000,%
Gitterman2004%
}.

The main aim of this paper is to study Ornstein-Uhlenbeck process with random damping
$\mu(t)$ based on a generalization of equation 
(\ref{eq:introduction_0040}):
\begin{align}
  \frac{dx(t)}{dt} & = -\mu(t) x + \zeta(t),
\label{eq:introduction_0041}
\end{align}
(\ref{eq:introduction_0041}) can be regarded as the overdamped limit of a special case of (\ref{eq:introduction_0020})
with $m(t)$ and $\gamma(t)$ equal to positive constants. 
Equivalently, it is the Langevin equation for the position $x(t)$ of the Brownian motion in a harmonic oscillator-like potential
$V(x)=\mu(t)x^2/2$.
We shall consider equation (\ref{eq:introduction_0041}) with fluctuating damping due to two kinds of multiplicative noise, 
namely the dichotomous noise and fractional Gaussian noise.
Note that random damping with fractional Gaussian noise has not been studied previously, 
though the case of dichotomous noise has been considered based on equation 
(\ref{eq:introduction_0011})
instead of 
(\ref{eq:introduction_0041}).
Explicit solutions of the resulting stochastic processes will be obtained with the full expressions of the mean and covariance of 
the associated process calculated to facilitate the study of their asymptotic behavior. We also study the stability of these solutions.


\section{Ornstein-Uhlenbeck Process with Multiplicative Noise}
In this section we consider the Ornstein-Uhlenbeck process obtained as the solution to the following Langevin equation with random damping or friction:
\begin{align}
  \frac{dx(t)}{dt} & = - \mu(t)x(t) +\chi(t),
\label{eq:OUMN_0010}
\end{align}
where $\mu(t)$ and $\chi(t)$ are random noises yet to be specified. For the ordinary Ornstein-Uhlenbeck
process, $\mu(t)$ is just a constant and $\chi(t)$ is given by $\eta(t)$ 
the usual Gaussian white noise defined by (\ref{eq:introduction_0012}).
Let
\begin{align}
  W(t) & = \int_0^t\mu(u)du,
\label{eq:OUMN_0030}
\end{align}
and
\begin{align}
  W(t|s) & = W(t) - W(s) = \int_s^t\mu(u)du .
\label{eq:OUMN_0040}
\end{align}
The solution of (\ref{eq:OUMN_0010}) can be written as
\begin{align}
  x(t) & = x_\circ G(t) + \int_0^tG(t|u)\chi(u)du,
\label{eq:OUMN_0050}
\end{align}
with
\begin{align}
  G(t) & = G(t|0)=e^{-W(t)},
\label{eq:OUMN_0060}
\end{align}
and
\begin{align}
  G(t|u) & = e^{-W(t|u)}.
\label{eq:OUMN_0070}
\end{align}

The mean and covariance of the process are respectively given by
\begin{align}
  \overline{x}(t) & = \bigl<x(t)\bigr> = x_\circ\bigl<G(t)\bigr> + \int_0^t \bigl<G(t|u)\bigr> \bigl<\chi(t)\bigr>, 
\label{eq:OUMN_0080} 
\end{align}
and
\begin{align}
K(t,s) & = \Bigl<\bigl[x(t) - \overline{x}(t)\bigr]\bigl[x(s) - \overline{x}(s)\bigr]\Bigr> \nonumber \\
       & = x_\circ^2 \Bigl[\bigl<G(t)G(s) - \overline{G}(t)\overline{G}(s)\bigr>\Bigr]
         + \int_0^{\min(t,s)} \bigl<G(t|u)G(s|u)\bigr>du.
\label{eq:OUMN_0090}
\end{align}
For convenience we denote the first and second terms in (\ref{eq:OUMN_0090}) by
$K_0(t,s)$ and $K_1(t,s)$.

In the subsequent sections we shall study two types of random damping based on the two
different multiplicative noise, namely dichotomous noise and fractional Gaussian noise. The case
of oscillator with fluctuating damping term given by dichotomous noise has been quite well-studied
\cite{Turelli1977,JasnowGerjuoy1975,%
GrueOeksendal1997,LuczkaTalknerHaenggi2000,%
Gitterman2004,Gitterman2005,LeprovostAumaitreMallick2006,%
MendezHorsthemkeMestresEtAl2011%
} 
(see also references given in \cite{Gitterman14}).
However, most of the studies do not calculate the covariance of the process 
(i.e. the solution of (\ref{eq:introduction_0010})), 
only mean and variance are calculated using the procedure of Shapiro-Loginov 
\cite{ShapiroLoginov1978}.
In this paper, the covariance of the process with dichotomous noise as random damping is calculated explicitly. 
Random damping in the form of fractional Gaussian noise so far has not been considered. 
Again, the covariance of the process with fractional Gaussian noise as damping term is computed explicitly. 
The effect of the multiplicative noise on stability of the process is considered.


\section{Ornstein-Uhlenbeck Process with Multiplicative Dichotomous Noise}
\label{sec:OUDin}

Denote by $\xi(t)$ the dichotomous (or telegraphic) process 
\cite{Kampen1992}
 which can be defined as
 \begin{align}
   \xi(t) & = \epsilon\xi_\circ(-1)^{N(t)},
\label{eq:OUDiN_0010}
 \end{align}
where $\epsilon < 1$, is a random variable which takes values ${\pm}1$, and $N(t)$ is a Poisson process with average rate $\lambda$.
The mean of the process is
\begin{align}
  \bigl<\xi(t)\bigr> & = \epsilon \bigl<\xi_\circ\bigr>\Bigl<(-1)^{N(t)}\Bigr>.
\label{eq:OUDiN_0020}
\end{align}
For $ t > s$, its covariance is then given by
\begin{align}
  \bigl<\xi(t)\xi(s)\bigr> & = \epsilon^2 e^{-2\lambda(t-s)}.
\label{eq:OUDiN_0030}
\end{align}
The solution to (\ref{eq:OUMN_0010}) with dichotomous noise as damping and 
$\chi(t)$ is the Gaussian white noise $\eta(t)$ is given by
(\ref{eq:OUMN_0050}) with
\begin{align}
  G(t|s) & = e^{-\int_s^t \xi(u)du}.
\end{align}

One can express $G(t)  = G(t|0)$ as a series:
\begin{align}
  G(t) & = \sum_{n=0}^\infty (-\epsilon)^n g_n(t),
\label{eq:OUDiN_0040}
\end{align}
with
\begin{align}
  g_n(t) & = \int_0^t dt_n \cdots \int_0^{t_2}dt_1 (-1)^{\sum_{k=1}^n N(t_k)}.
\label{eq:OUDiN_0050}
\end{align}
The series (\ref{eq:OUDiN_0040}) can be rewritten as the sum of even and odd powers of $\epsilon$:
\begin{align}
  G(t) & = \sum_{n=0}^\infty (-\epsilon)^{2n}g_{2n}(t) + \sum_{n=0}^\infty (-\epsilon)^{2n+1}g_{2n+1}(t) \nonumber \\
       & = U(t) - V(t),
\end{align}
where
\begin{align}
  U(t) & = \sum_{n=0}^\infty \epsilon^{2n}g_{2n}(t) &
       & \text{and} &
  V(t) & = \sum_{n=0}^\infty \epsilon^{2n+1}g_{2n+1}(t) .
\label{eq:OUDiN_0060}
\end{align}

It is straight forward to show that $\overline{g}_0(t) =1$, $\overline{g}_1(t) = \frac{1-e^{-2\lambda{t}}}{2\lambda}$,
and for $n \geq 2$,
\begin{align}
  \overline{g}_n(t) & = \int_0^t f(t-u)\overline{g}_{n-2}(u)du.
\end{align}
Denote $\overline{g}_1(t)$ by $f(t)$.
The mean value of $U(t)$, $V(t)$ and $G(t)$ are then given by
\begin{subequations}
\label{eq:OUDiN_0070}
  \begin{align}
    \overline{U}(t) & = 1 + \epsilon^2 \int_0^t f(u)\overline{U}(u)du,\label{eq:OUDiN_0070a} \\
    \overline{V}(t) & = \epsilon f(t) + \epsilon^2 \int_0^t f(u)\overline{V}(u)du, \label{eq:OUDiN_0070b}
  \end{align}
and
\begin{align}
      \overline{G}(t) & = 1 - \epsilon f(t) + \epsilon^2 \int_0^t f(u)\overline{G}(u)du. 
\label{eq:OUDiN_0070c}
\end{align}
\end{subequations}
We remark that the above average values are conditional averages with initial random variable $\xi_\circ = 1$.
By averaging over $\xi_\circ = \pm 1$  with equal probability, 
all odd terms with respect to $\epsilon$ vanish, one gets
$\overline{G}=\overline{U}$.

(\ref{eq:OUDiN_0070}) can be evaluated with the help of Laplace transform to give
\begin{subequations}
\label{eq:OUDiN_0080}
  \begin{align}
    \overline{U}(t) & = \frac{1}{2\Lambda}\Bigl[-\beta e^{\alpha{t}} + \alpha e^{\beta{t}}\Bigr], \label{eq:OUDiN_0080a}\\
    \overline{V}(t) & = \frac{\epsilon}{2\Lambda}\Bigl[ e^{\alpha{t}} - e^{\beta{t}}\Bigr], \label{eq:OUDiN_0080b} \\
    \overline{G}(t) & = \frac{1}{2\Lambda}\Bigl[-(\beta+\epsilon) e^{\alpha{t}} + (\alpha+\epsilon) e^{\beta{t}}\Bigr], \label{eq:OUDiN_0080c}
  \end{align}
\end{subequations}
where
\begin{align}
  \Lambda & = \sqrt{\lambda^2+\epsilon^2}, &
   \alpha & = -\lambda + \Lambda, &
    \beta & = -\lambda - \Lambda.
\label{eq:OUDiN_0090}
\end{align}
The mean of the process is then given by
\begin{align}
  \overline{x}(t) & = \overline{U}(t)x_\circ .
\label{eq:OUDiN_0100}
\end{align}

In order to determine the covariance 
(\ref{eq:OUMN_0090})
of the process, it is necessary first to calculate
$\Bigl<G(t|u)G(s|u)\Bigr>$,
which can be rewritten as
\begin{align}
  \Bigl<G(t|u)G(s|u)\Bigr> & = \Bigl<G(t|s)G^2(s|u)\Bigr> \nonumber\\
                           & = \Bigl<U(t-s)\Bigr>\Bigl<G^2(s|u)\Bigr> - \Bigl<V(t-s)\Bigr>\frac{1}{(-2\epsilon)}\frac{d}{ds}\Bigl<G^2(s|u)\Bigr>.
\label{eq:OUDiN_0110}
\end{align}
(See the appendix
for more details.)
By noting that $G^2(\cdot)$ has the same expression as $G(\cdot)$ with $\epsilon$ replaced by $2\epsilon$,
and using a similar argument as given in 
\ref{sec:CalDiNGG}, one can obtain
\begin{align}
  \overline{G^2}(s|u) = \overline{U^2}(s-u) - \overline{V^2}(s-u)e^{-2\lambda{u}},  
\label{eq:OUDiN_0120}
\end{align}
and
\begin{align}
  \frac{d}{ds}\overline{U^2}(s-u) & = 2\epsilon \overline{V^2}(s-u).
\label{eq:OUDiN_0130}
\end{align}
Now taking the average over initial random variable $\xi_\circ = \pm{1}$,
noting that only even terms in $\epsilon$ remain, one thus gets
\begin{align}
  \Bigl<G(t|u)G(s|u)\Bigr> & = \overline{U}(t-s)\overline{U^2}(s-u) + \overline{V}(t-s)\overline{V^2}(s-u).
\label{eq:OUDiN_0140}
\end{align}
Using the above results, one finally obtains $K(t,s) = K_\circ(t,s) + K_1(t,s)$ with
\begin{subequations}
\label{eq:OUDiN_0150}
  \begin{align}
    K_\circ(t,s) & = x_\circ^2\Bigl[
                               \overline{U}(t-s)\overline{U^2}(s) + \overline{V}(t-s)\overline{V^2}(s) -\overline{U}(t)\overline{U}(s)
                            \Bigr], \label{eq:OUDiN_0150a} \\
    K_1(t,s)    & =  \overline{U}(t-s)\int_0^s \overline{U^2}(s-u)du + \overline{V}(t-s)\int_0^s \overline{V^2}(s-u)du.
\label{eq:OUDiN_0150b}
  \end{align}
\end{subequations}
Let $\Delta = t - s$ in (\ref{eq:OUDiN_0150}).   
By carrying out the integrations in (\ref{eq:OUDiN_0150b}), and expanding the sum of its argument
$\overline{U}(t) = \overline{U}(\Delta)\overline{U}(s) + \overline{V}(\Delta)\overline{V}(s)$
in (\ref{eq:OUDiN_0150a}), and rearranging the terms gives
\begin{subequations}
\label{eq:OUDiN_0160}
\begin{align}
  K_\circ(s+\Delta,s) & = x_\circ^2 
                        \Bigl[
                        \overline{U}(\Delta)\overline{U^2}(s) + \overline{V}(\Delta)\overline{V^2}(s) 
                        -\bigl(\overline{U}(\Delta)\overline{U}(s) + \overline{V}(\Delta)\overline{V}(s)\bigr)\overline{U}(s)
                        \Bigr], \label{eq:OUDiN_0160a}\\
 K_1(s+\Delta,s) & = \overline{U}(\Delta)
                   \biggl[
                   \frac{\lambda}{2\epsilon^2}\Bigl(\overline{U^2}(s) -\overline{U^2}(0)\Bigr)
                   + \frac{1}{2\epsilon}\overline{V^2}(s)
                   \biggr]
                   + \frac{1}{2\epsilon}\overline{V}(\Delta)\Bigl(\overline{U^2}(s) -\overline{U^2}(0)\Bigr).
\label{eq:OUDiN_0160b}
\end{align}  
\end{subequations}

Now we want to consider the properties of the mean and covariance, in particular their asymptotic behavior. 
For $\epsilon=0$, one get $\Lambda=\lambda$, $\alpha=0$, and $\beta = -2\lambda$, which immediately imply
$\overline{U}(t)=1$ and $\overline{V}(t) = 0$.
Hence one gets the mean $\overline{x}(t) = x_\circ$.
By taking the $\epsilon \to 0$ limit in (\ref{eq:OUDiN_0160}) results in $K(t,s) = s$, 
a result consistent with the covariance of Brownian motion.

Next, we consider the small-time and long-time asymptotic properties of the covariance.
In the $t \to 0$ limit, up to first order of $t$, one has $\overline{U}(t) = 1$ and $\overline{V}(t) = \epsilon{t}$.
Thus, for small $s$ and finite $\Delta = t-s$,
\begin{align}
  K(s+\Delta,s) & = \Bigl[\overline{U}(\Delta) + \epsilon{x_\circ^2}\overline{V}(\Delta)\Bigr]s.
\label{eq:OUDiN_0170}
\end{align}
For $t \to \infty$, one can neglect $e^{\beta{t}}$ and just keep $e^{\alpha{t}}$ since $\alpha > \beta$.
This gives 
\begin{align}
\overline{U}(t) & = \frac{-\beta}{2\Lambda}e^{\alpha{t}} &
                & \text{and} &
\overline{V}(t) & = \frac{\epsilon}{2\Lambda}e^{\alpha{t}}.
\label{eq:OUDiN_0180}
\end{align}
The mean is then given by an exponential increasing function:
\begin{align}
  \overline{x}(t) & = x_\circ \frac{-\beta}{2\Lambda}e^{\alpha{t}}.
\label{eq:OUDiN_0190}
\end{align}

The long-time limit of the covariance is more delicate. 
We have the following asymptotic expressions:
\begin{subequations}
\label{eq:OUDiN_0200}
\begin{align}
  K_\circ(s+\Delta,s) & \sim x_\circ^2 
                        \Biggl[
                        \bigg(
                        \overline{U}(\Delta) + \frac{2\epsilon}{-\beta_2}\overline{V}(\Delta)
                        \biggr)
                        \biggl(
                        \frac{-\beta_2}{2\Lambda_2}
                        \biggr)
                        e^{\alpha_2{s}}
                        -\biggl(
                        \overline{U}(\Delta) + \frac{\epsilon}{-\beta}\overline{V}(\Delta)
                        \biggr)
                        \biggl(
                        \frac{\beta^2}{4\Lambda}
                        \biggr)
                        e^{2\alpha{s}}
                        \Biggr],  \label{eq:OUDiN_0200a} \\
K_1(s+\Delta,s) & \sim \Biggl[
                  \overline{U}(\Delta)
                  \biggl(
                  \lambda + \frac{2\epsilon^2}{-\beta_2}              
                  \biggr)
                  + \epsilon\overline{V}(\Delta)
                  \Biggr]
                  \biggl(
                  \frac{-\beta_2}{4\epsilon^2\Lambda_2}
                  \biggr)
                  \Bigl(
                  e^{\alpha_2{s}} - 1
                  \Bigr) .
\label{eq:OUDiN_0200b}
  \end{align}
\end{subequations}
For the reason that $\alpha_2 > 2\alpha$, but very small (for any fixed $t$), the two exponential terms in
(\ref{eq:OUDiN_0200a}) cannot be ignored.
However, for very large $s$ we could said that it behaves asymptotically as $e^{\alpha_2{s}}$,
that is it depends on only the first term.
For (\ref{eq:OUDiN_0200b}) it varies asymptotically as $e^{\alpha_2{s}} -1$,
and it is obvious in the asymptotic sense it behaves like $e^{\alpha_2{s}}$.
However, the expression not only has this asymptotic behavior, so is its tangent at infinity. 
Moreover, it takes care of the term $1/\epsilon^2$, that is it cancels this divergence. 
Thus for fixed $\epsilon > 0$, one can say that the covariance increases asymptotically as $e^{\alpha_2{s}}$.

Next, we want to consider a more conventional way of introducing the random damping term. 
If instead of the dichotomous random damping $\mu(t)$,
one modifies it as the sum $\mu(t) = \mu_\circ + \epsilon\xi(t)$,
where $\xi(t)$ is the dichotomous noise as defined earlier. 
In other words, the dichotomous noise is now introduced as a perturbation term to the original constant damping $\mu_\circ$.
Let the resulting process be denoted by $x_{+}(t)$.
In this case it is straight forward to obtain the $G_{+}$ corresponding to $x_{+}(t)$:
\begin{align}
  G_{+}(t|s) & = e^{-\mu_\circ(t-s)}G(t|s),
\label{eq:OUDiN_0210}
\end{align}
where $G(t|s)$ is the same as given before by \mbox{(\ref{eq:OUMN_0070})}.
Similarly, the associated mean is given by
\begin{align}
  \overline{x}_{+}(t) & = x_\circ\overline{U}_{+}(t) = x_\circ e^{-\mu_\circ{t}}\overline{U}(t)
                       = x_\circ e^{-\mu_\circ{t}}\Bigl[-\beta e^{\alpha{t}} + \alpha e^{\beta{t}}\Bigr].
\label{eq:OUDiN_0220}
\end{align}
Clearly, as $t \to \infty$ the mean $\overline{x}_{+}(t)$ converges to zero.

\begin{figure}[t!]
  \centering
\includegraphics[width=0.8\textwidth]{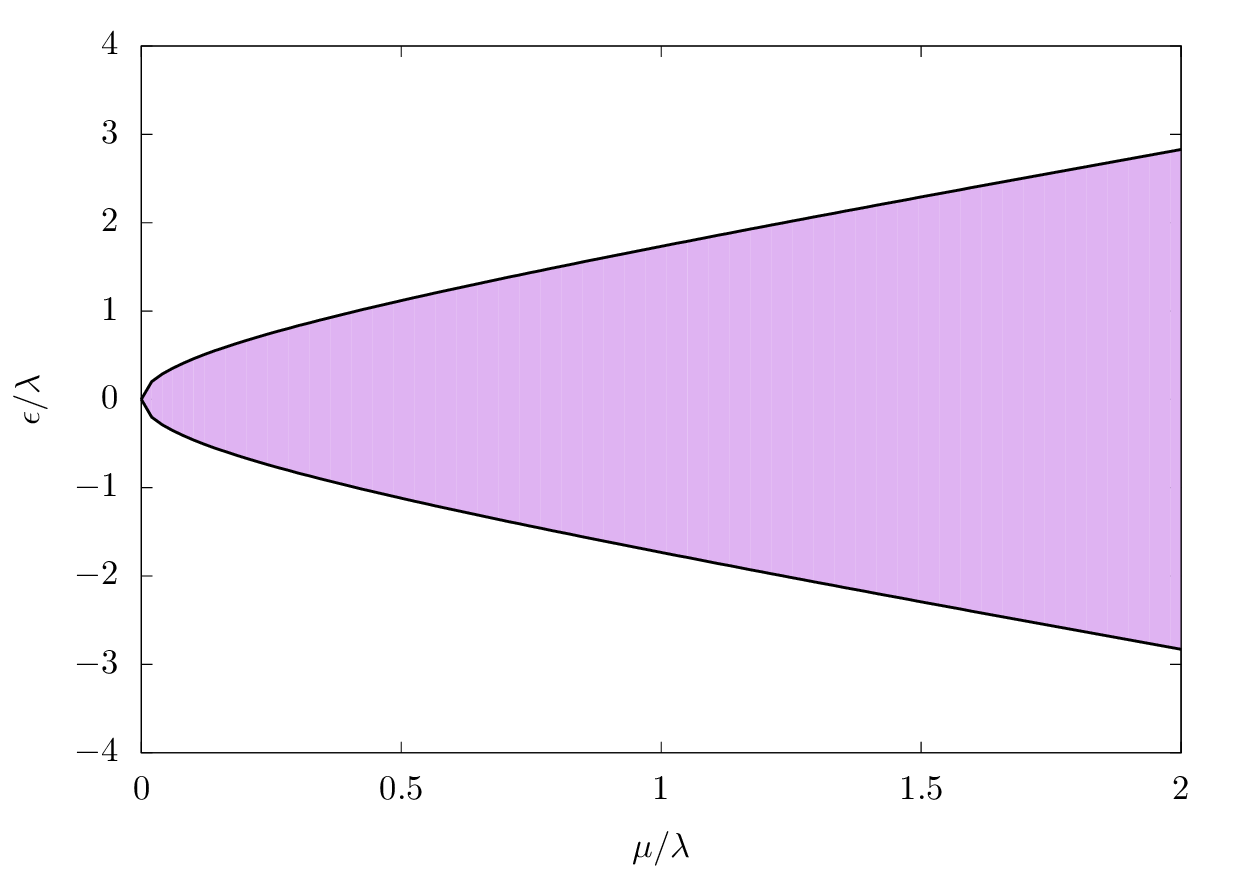}
\caption{The process is mean-stable in the shaded region}
\label{fig:DINP_0011} 
\end{figure}

Note that the case $\mu_\circ = \alpha$ is the critical value as $\mu_\circ < \alpha$ the mean is not stable, and $\mu_\circ \geq \alpha$ it is stable. 
By using (\ref{eq:OUDiN_0090})  the critical point can be expressed explicitly as
\begin{align}
  \mu_\circ & = - \lambda + \sqrt{\lambda^2 +\epsilon^2}
\label{eq:OUDiN_0221}
\end{align}
Dividing (\ref{eq:OUDiN_0221}) by $\lambda$, and rearranging and then taking square of the resulting expression gives
\begin{align}
  \biggl(\frac{\mu_\circ}{\lambda}+1\biggr)^2 - \biggl(\frac{\epsilon}{\lambda}\biggr)^2 & = 1,
\label{eq:OUDiN_0222}
\end{align}
which is the equation of a hyperbola. The stability condition becomes 
$\Bigl(\frac{\mu_\circ}{\lambda}+1\Bigr)^2 - \Bigl(\frac{\epsilon}{\lambda}\Bigr)^2 \geq 1$.
Figure \ref{fig:DINP_0011} shows only the right part of the hyperbola (\ref{eq:OUDiN_0222}). 
The values inside the shaded area give stable mean.

\begin{figure}[t!]
  \centering
\includegraphics[width=0.8\textwidth]{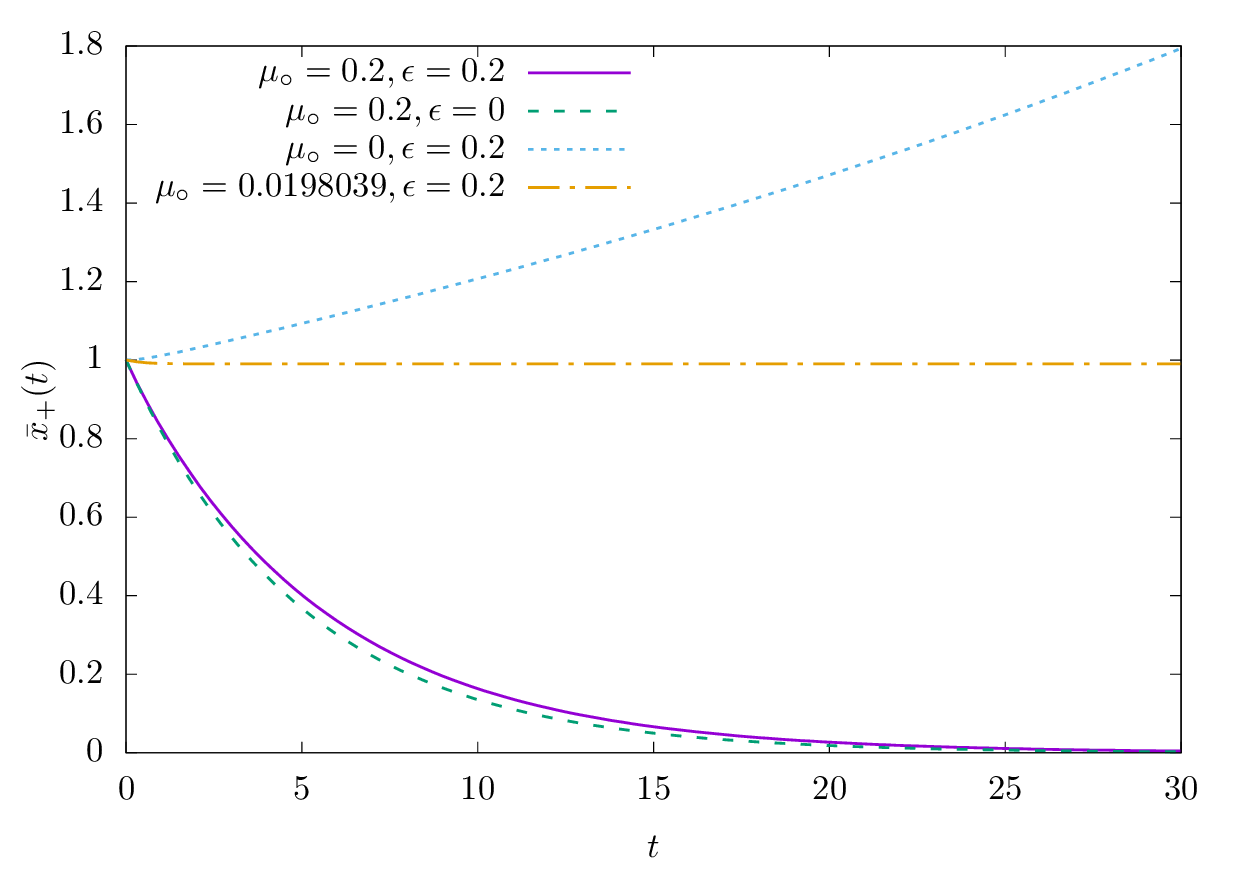}
\caption{Mean of the process $x_{+}(t)$ with different values of $\mu_\circ$ and $\epsilon$ }
\label{fig:DINP_0010} 
\end{figure}

Note that at the critical value, we have
\begin{align}
  \overline{x}_{+}(t) = x_\circ \frac{1}{2\Lambda}\Bigl[-\beta + \alpha e^{-2\Lambda{t}}\Bigr].
\label{eq:OUDiN_0230}
\end{align} 

Here we remark that the process is stable in the mean at the critical point even though the value 
$\overline{x}_{+}(t)$ approaches asymptotically to a constant $\overline{x}_{+}(t) - x_\circ \frac{\lambda + \Lambda}{2\Lambda}$
which is a slightly lower than $x_\circ$.
However, if we use the Lyapunov stability condition (to be discussed in section \ref{sec:stability}),
then it is not stable at the critical point. 
Figure \ref{fig:DINP_0010} depicts how the mean $\overline{x}_+$ changes with time for different values of $\mu_\circ$ and $\epsilon$.

As for the covariance, one has
\begin{align}
  K_{\circ+}(t,s) & = x_\circ^2\Bigl[
                    \overline{U}_{+}(t-s)\overline{U^2}_{+}(s) + \overline{V}_{+}(t-s)\overline{V^2}_{+}(s)
                    - \overline{U}_{+}(t)\overline{U}_{+}(s)
                    \Bigl].
\label{eq:OUDiN_0240}
\end{align}
The second component of the covariance function is slightly more involved. 
Consider the expression analogous to (\ref{eq:OUDiN_0150b}), and after integration one gets
\begin{multline}
  K_{1+}(t,s)  = \frac{\overline{U}_{+}(\Delta)}{2\Lambda_2}
                       \biggl[
                       \frac{-\beta_2}{2\mu_\circ - \alpha_2}\Bigl(1-e^{-(2\mu_\circ -\alpha_2)s}\Bigr) 
                       +
                       \frac{\alpha_2}{2\mu_\circ - \beta_2}\Bigl(1-e^{-(2\mu_\circ -\beta_2)s}\Bigr)
                       \biggr] \\
                       +
                       \frac{2\epsilon \overline{V}_{+}(\Delta)}{2\Lambda_2}
                       \biggl[
                       \frac{1}{2\mu_\circ - \alpha_2}\Bigl(1-e^{-(2\mu_\circ -\alpha_2)s}\Bigr) 
                       -
                       \frac{1}{2\mu_\circ - \beta_2}\Bigl(1-e^{-(2\mu_\circ -\beta_2)s}\Bigr)
                       \biggr].
\label{eq:OUDiN_0250}
\end{multline}
Here, 
\begin{align}
    \Lambda_2 & = \sqrt{\lambda^2+4\epsilon^2}, &
   \alpha_2 & = -\lambda + \Lambda_2, &
    \beta_2 & = -\lambda - \Lambda_2.
\label{eq:OUDiN_0251}
\end{align}

\begin{figure}[t!]
  \centering
  \includegraphics[width=0.8\textwidth]{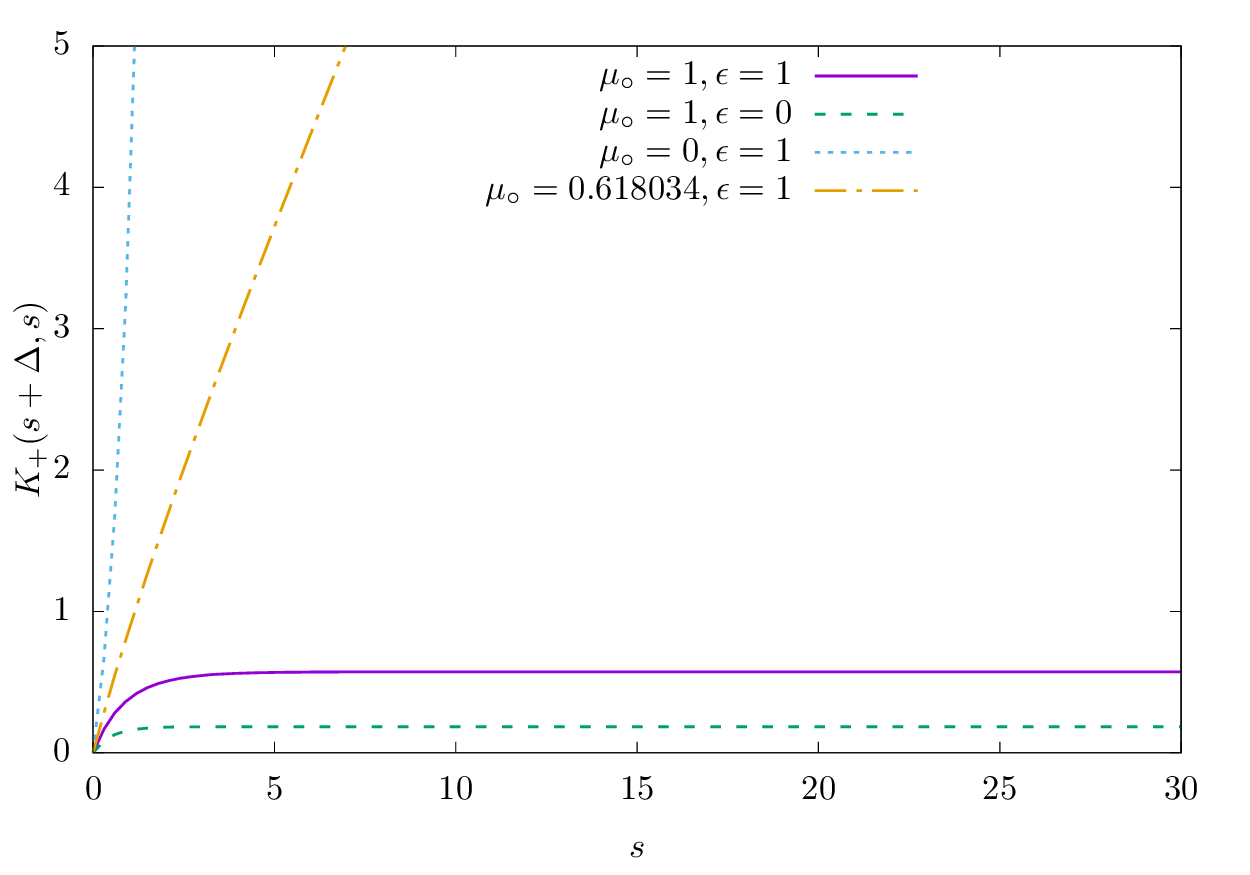}
\caption{Covariance of the process $x_{+}(t)$ with different values of $\mu_\circ$ and $\epsilon$ for fixed $\Delta=1$}
\label{fig:DINP_0020}
\end{figure}

In the case of covariance, the critical value for it to be bounded is $\mu_\circ = \alpha_2/2$.
One can obtain a hyperbola equation similar to that of (\ref{eq:OUDiN_0222}):
\begin{align}
  \biggl(\frac{2\mu_\circ}{\lambda}+1\biggr)^2 - \biggl(\frac{2\epsilon}{\lambda}\biggr)^2 & = 1
\label{eq:OUDiN_0223}
\end{align}

For $\mu_\circ > \alpha_2/2$ and finite $\delta = t-s$, $s \to \infty$, one obtains
\begin{align}
  K_{1+}(s+\Delta,s) &  = \frac{\overline{U}_{+}(\Delta)}{2\Lambda_2}
                       \biggl[
                       \frac{-\beta_2}{2\mu_\circ - \alpha_2}
                       +
                       \frac{\alpha_2}{2\mu_\circ - \beta_2}
                       \biggr] \nonumber \\
                    & \qquad   +
                       \frac{2\epsilon \overline{V}_{+}(\Delta)}{2\Lambda_2}
                       \biggl[
                       \frac{1}{2\mu_\circ - \alpha_2}
                       -
                       \frac{1}{2\mu_\circ - \beta_2}
                       \biggr].
\label{eq:OUDiN_0260}
\end{align}
So the process is stationary in the large-$t$ limit. 
It converges to a bounded value independent of the initial point. 
Figure \mbox{\ref{fig:DINP_0020}} shows the covariance corresponds to various values  
of $\mu_\circ$ and $\epsilon=1$ for fixed $\Delta =1$.
For $\mu_\circ \leq \alpha_2/2$  the covariance is 
Finally we note that for both above cases it is clear that $\frac{\epsilon}{\mu} < 1$,
which justifies the use of the term perturbative.


\section{Ornstein-Uhlenbeck Process with Multiplicative Fractional Gaussian Noise}
\label{sec:OUfGN}
In this section we want to consider the random damping $\mu(t)$ given by fractional Gaussian noise $V(t)$, 
that is the noise obtained from the generalized derivative (in the sense of generalized function) of the fractional Brownian motion, 
just like white noise is the generalized derivative of Brownian motion 
\cite{MandelbrotNess1968}.
The fractional Brownian motion $W(t)$ is defined as a Gaussian process with mean zero and the following covariance 
\cite{MandelbrotNess1968,Samorodnitsky1994}:
\begin{align}
  C(t,s) & = \bigl<W(t)W(s)\bigr> - \bigl<W(t)\bigr>\bigl<W(s)\bigr>  = \frac{\sigma_H^2}{2}\Bigl[|t|^{2H} + |s|^{2H} - |t-s|^{2H}\Bigr],
\label{eq:OUfGN_0010}
\end{align}
where $0 < H < 1$ is the Hurst index. Note that the form of the coefficient $\sigma_H^2$ depends 
on the constant coefficient in the definition of $W(t)$. 
For examples, it can be normalized to be unity, or it can also be given explicitly as
\begin{align}
  \sigma_H^2 & = \frac{\Gamma(2-2H)\cos(\pi{H})}{\pi{H}(1-2H)}.
\label{eq:OUfGN_0020}
\end{align}
For simplicity we assume $\sigma_H^2$ to be unity. 
Fractional Gaussian noise $V(t)$ can considered as the generalized derivative of $W(t)$, 
is defined as a Gaussian process with zero mean and covariance given by
\begin{align}
  \bigl<V(t)V(t+\tau)\bigr> & = H(2H-1)|\tau|^{2H-2|} .
\label{eq:OUfGN_0030}
\end{align}

Now by using the same notation $G(\cdot,\cdot)$ as for the previous sections, one gets
\begin{align}
  \bigl<G(t|s)\bigr> & = e^{\frac{1}{2}|t-s|^{2H}} .
\label{eq:OUfGN_0040}
\end{align}
\begin{align}    
  & \bigl<G(t|u)G(s|v)\bigr>  = \exp\biggl\{\frac{1}{2}\Big<\big[W(t)-W(u)+W(s)-W(v)\bigr]^2\Big>\biggr\} \nonumber                \\
  &  = \exp\biggl\{
    \frac{1}{2}\Big[
    |t-u|^{2H} - |t - s|^{2H} + |t - v|^{2H} + |s - u|^{2H} + |s - v|^{2H} - |u - v|^{2H} 
    \Bigr]
         \biggr\} . \label{eq:OUfGN_0060}
\end{align}
The Ornstein-Uhlenbeck process with fractional Gaussian noise as the damping term, denoted by $x_g$, has the mean given by
\begin{align}
  \overline{x}_g(t) & = x_\circ\overline{G}(t) = x_\circ e^{\frac{1}{2}|t|^{2H}}.
\label{eq:OUfGN_0070}
\end{align}

The covariance can be split into two parts as it has been done previously:
\begin{subequations}
\label{eq:OUfGN_0080}
\begin{align}
  K_{g\circ}(t,s) & = x_\circ^2 \biggl[
                        e^{\frac{1}{2}\bigl[2|t|^{2H} + 2|s|^{2H} - |t-s|^{2H}\bigr]}                                 
                        - e^{\frac{1}{2}\bigl[|t|^{2H} + |s|^{2H}\bigr]}                                 
                        \biggr] , \label{eq:OUfGN_0080a} \\
  K_{g1}(t,s) & = \int_0^s e^{\frac{1}{2}\bigl[2|t-u|^{2H} + 2|s-u|^{2H} - |t-s|^{2H}\bigr]} du . \label{eq:OUfGN_0080b}
\end{align}  
\end{subequations}

The integration in (\ref{eq:OUfGN_0080b}) can be evaluated by substitution $u^\prime = s -u$, which gives
\begin{subequations}
\begin{align}
  K_{g1}(t,s) & = e^{-\frac{1}{2}|t-s|^{2H}}\int_0^s e^{\frac{1}{2}\bigl[2|t-s+u|^{2H} + 2|u|^{2H}\bigr]} du  \\
             & = e^{-\frac{1}{2}|t-s|^{2H}} \sum_{m=0}^\infty\sum_{n=0}^\infty \frac{1}{m!n!}\int_0^s |t-s+u|^{2mH}|u|^{2nH} du.
\label{eq:OUfGN_0090}
\end{align}
\end{subequations}
(\ref{eq:OUfGN_0080b}) can also be written in the following form:
\begin{subequations}
\begin{align}
  K_{g1}(t,s) & = se^{-\frac{1}{2}|t-s|^{2H}}\int_0^1 e^{\frac{1}{2}\bigl[2t^{2H}|1-(s/t)u|^{2H} + 2s^{2H}|1-u|^{2H}\bigr]} du \\
             & = e^{-\frac{1}{2}|t-s|^{2H}} \sum_{m,n=0}^\infty \frac{t^{2mH}s^{2nH+1}}{m!n!}\int_0^1\bigl|1-(s/t)u\bigr|^{2mH}\bigl|1-u\bigr|^{2nH}du .
\label{eq:OUfGN_0100}
\end{align}  
\end{subequations}
From the integral table \cite[p.317,\#3.197.3]{GradshteynRyzhik1980}
\begin{align}
  \int_9^1 x^{\lambda-1}(1-x)^{\mu-1}(1-\beta{x})^{-\nu} dx & = B(\lambda,\mu){_2F_1}(\nu,\lambda,\lambda+\mu;\beta) \label{eq:OUfGN_0110}\\
                                                       & \quad [Re\lambda > 0, Re\mu > 0, |\beta| < 1] , \nonumber 
\end{align}
and by letting
\begin{align}
  \lambda & = 1, &
  \mu     & = 2nH+1, &
  \beta   & = s/t,
\label{eq:OUfGN_0120}
\end{align}
one gets
\begin{align}
  K_{g1}(t,s) & = e^{-\frac{1}{2}|t-s|^{2H}} \sum_{m,n=0}^\infty \frac{t^{2mH}s^{2nH+1}}{m!n!}B(1,2nH+1){_2F_1}(-2mH,1,2nH+2;s/t) .
\label{eq:OUfGN_0130}
\end{align}

Let the variance be denoted by 
\begin{align*}
  S_g(t) & = K_g(t,t) = K_{g\circ}(t,t) + K_{g1}(t,t) = S_{g\circ}(t) + S_{g1}(t).
\end{align*}
From (\ref{eq:OUfGN_0080a}), we have
\begin{align}
  S_{g\circ}(t)  & = x_\circ^2e^{|t|^{2H}}\Bigl[e^{|t|^{2H}} -1\Bigr].
\label{eq:OUfGN_0140}
\end{align}
One can get $S_{g1}(t)$ from (\ref{eq:OUfGN_0130}) by letting $s/t=1$.  
A more transparent expression is obtained by using (\ref{eq:OUfGN_0080b}) with $s=t$:
\begin{align}
 S_{g1}(t) & = \int_0^t e^{2|t-u|^{2H}} du = \int_0^t e^{2|u|^{2H}} du,
\label{eq:OUfGN_0150}
\end{align}
which can be evaluated by direct expansion:
\begin{align}
  S_{g1}(t) & = \sum_{n=1}^\infty (2)^n \frac{t^{2nH+1}}{n!(2nH+1)} .
\label{eq:OUfGN_0160}
\end{align}

For special case $2H=1/q$ for $q = 1,2, \cdots$, it has a closed expression
\begin{align}
  S_{g1}(t) & = q\left(\frac{1}{2}\right)^q \int_0^{2t^{1/q}} u^{q-1}e^u du \nonumber \\
           & = q\left(\frac{1}{2}\right)^q 
             \biggl[
             (-1)^q(q-1)! - e^{2t^{1/q}}\sum_{k=0}^{q-1} \frac{(q-1)!}{k!}(-1)^{q-k}2^kt^{k/q}
             \biggr] .
\label{eq:OUfGN_0170}
\end{align}
When $q=2$ (or $H=1/4$) one gets
\begin{align}
  S_{g1}(t) & = \frac{1}{2}\Bigl[1-e^{2\sqrt{t}} + 2\sqrt{t}e^{2\sqrt{t}}\Bigr].
\label{eq:OUfGN_0180}
\end{align}
The short and long time behaviors of (\ref{eq:OUfGN_0180}) are given by
$S_{g1}(t) \sim t$ and $S_{g1}(t) \sim \sqrt{t}e^{2\sqrt{t}}$.
However, the behaviors for $S_{g\circ}(t)$ is $\sim x_\circ^2\sqrt{t}$ and $\sim x_\circ e^{2\sqrt{t}}$, coresponding.
Therefore the short and long time total variance varies as $S_g(t)$ is $\sim x_\circ^2\sqrt{t}$ and $\sim \sqrt{t}e^{2\sqrt{t}}$.

\begin{figure}[ht]
  \centering
  \includegraphics[width=0.8\textwidth]{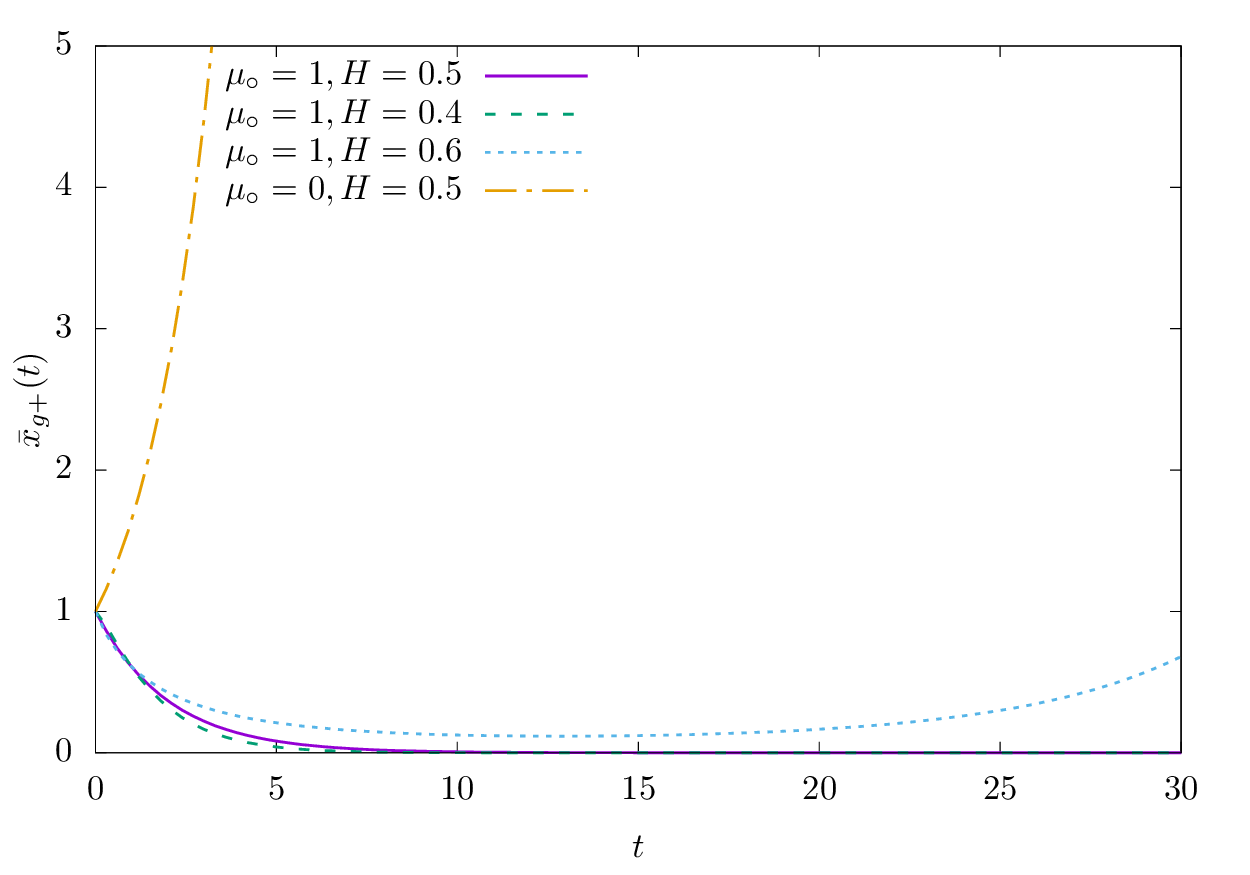}
  \caption{Time variation of mean $\overline{x}_{g+}(t)$ for different values of $\mu_\circ $ and $H$ with $\epsilon=1$.}
\label{fig:OUfGN_0030}
\end{figure}

\begin{figure}[ht]
  \centering
  \includegraphics[width=0.8\textwidth]{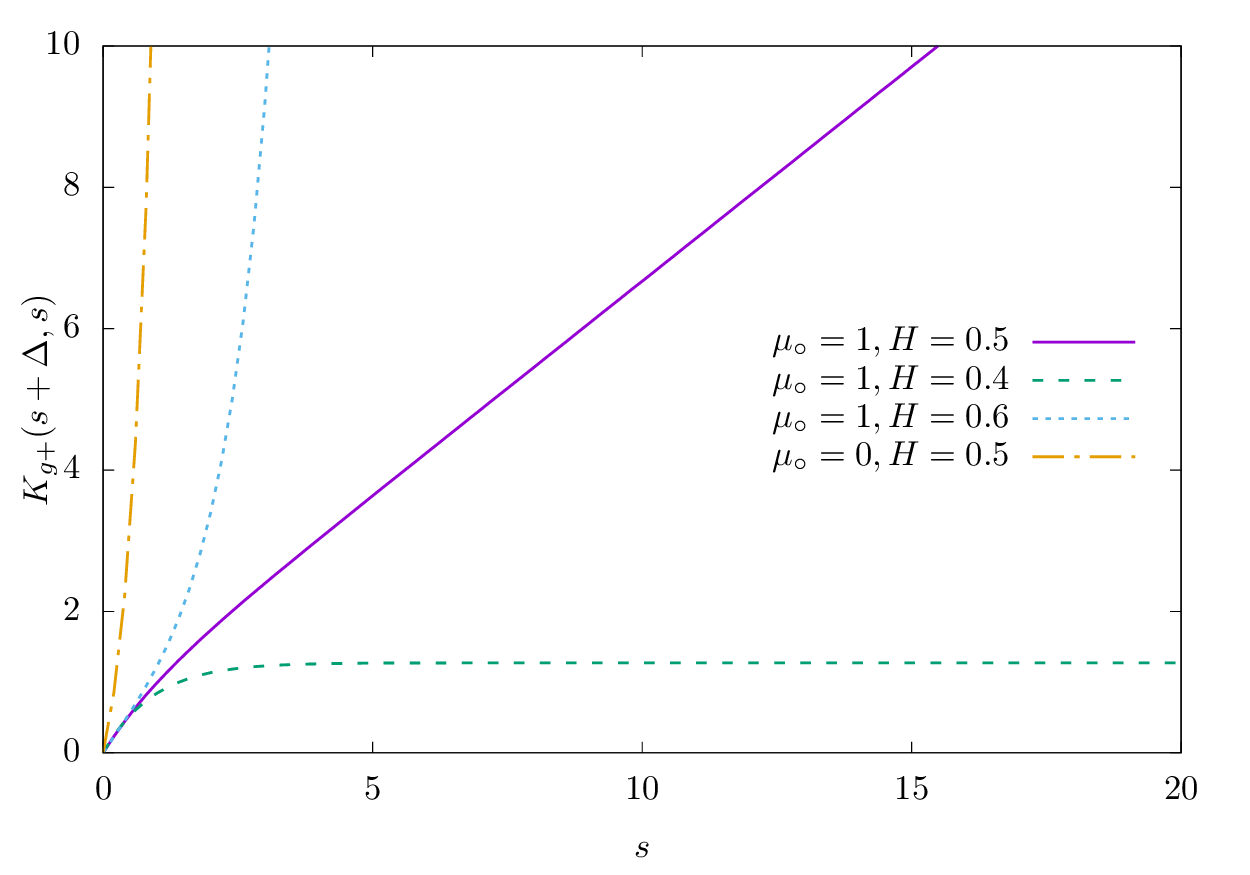}
\caption{Covariance of $x_{g+}(t)$ for different values of  $\mu_\circ$ and $H$ with $\epsilon = 1$ and $\Delta = 1$}
\label{fig:OUfGN_0040}
\end{figure}

Again, just like in the case of dichotomous noise, we can introduce the fractional Gaussian noise as a perturbation to the original damping so that the modified damping term becomes
$\mu(t) = \mu_\circ + \epsilon\xi(t)$.
The resulting process has the following mean and covariance:
\begin{align}
  \overline{x}_{g+}(t) & = x_\circ \overline{G}_+(t) = x_\circ e^{-\mu_\circ{t} + \frac{\epsilon^2}{2}|t|^{2H}}.
\label{eq:OUfGN_0181}
\end{align}
and
\begin{subequations}
\label{eq:OUfGN_0190}
  \begin{align}
 K_{g\circ}(t,s) & = x_\circ^2 e^{-\mu_\circ(t+s)} \biggl[
                        e^{\frac{\epsilon^2}{2}\bigl[2|t|^{2H} + 2|s|^{2H} - |t-s|^{2H}\bigr]}                                 
                        - e^{\frac{\epsilon^2}{2}\bigl[|t|^{2H} + |s|^{2H}\bigr]}                                 
                        \biggr] , \label{eq:OUfGN_0190a}\\
 K_{g1}(t,s) & = \int_0^s e^{-\mu_\circ(t+s -2u)+\frac{\epsilon^2}{2}\bigl[2|t-u|^{2H} + 2|s-u|^{2H} - |t-s|^{2H}\bigr]} du \nonumber \\
              & = e^{-\mu_\circ(t-s)+\frac{\epsilon^2}{2}|t-s|^{2H}}\int_0^s e^{-\mu_\circ(2u)+\frac{\epsilon^2}{2}\bigl[2|t-s+u|^{2H} + 2|u|^{2H}\bigr]} du \nonumber \\
              & = |t-s|e^{-\mu_\circ(t-s)-\frac{\epsilon^2}{2}|t-s|^{2H}}
                   \int_0^{s/|t-s|} e^{-2\mu_\circ{|t-s|u} +\epsilon^2|t-s|^{2H}\bigl[|1+u|^{2H} + |u|^{2H}\bigr]} du . \label{eq:OUfGN_0190b}
  \end{align}
\end{subequations}
and again we define $K_{g+}(t,s) = K_{g0+}(t,s) + K_{g1+}(t,s)$.
Figure \mbox{\ref{fig:OUfGN_0030}} and  \mbox{\ref{fig:OUfGN_0040}}
depict respectively the variation of mean in time for various values of $\mu_\circ$  and $H$ for $\epsilon =1$,
and the changes of covariance with different values of $\mu_\circ$ and $H$ for $\epsilon =1$ and $\Delta =1$.

The condition for $x_{g+}(t)$ to be stable in the mean and covariance is independent of $\epsilon$ as far as both
$\epsilon > 0$ and $\mu_\circ > 0$;
it depends only on the Hurst index. 
This can be easily seen by considering the expressions given by (\ref{eq:OUfGN_0181}) and (\ref{eq:OUfGN_0190}). 
The dominant term for $2H < 1$ is the exponential $e^{-\mu_\circ{t}}$ as all powers of $t^{2H}$  approach infinity. 
This case give stable mean and asymptotic stationary covariance. 
However, in the case $2H=1$ and exponential terms are linear in time. 
Thus the stability is determined by $\mu_\circ > \epsilon^2/2$.

From (\ref{eq:OUfGN_0190b}), it can be immediately notice that after limit $s \to \infty$, it becomes a Laplace transform.
Moreover, it is a function of the difference $t-s$, thus $K_{g1}$ is stationary.
The $s \to \infty$ limit \mbox{(\ref{eq:OUfGN_0190a})} vanishes.
Thus, it is a stationary as well.

\section{Stability}
\label{sec:stability}

In this section we want to discuss one important property of the Ornstein-Uhlenbeck process with random damping, 
namely its stability. 
Stability of a solution of differential equation has been well studied 
(see for example \cite[and reference therein]{AnishchenkoVadivasovaStrelkova2014}).
There exist many variants on the notions on stability such as uniform stability, exponential stability etc.  
The most widely discussed topic on this subject is the Lyapunov stability which we shall consider later on. 
For the general non-linear equation
\begin{align}
  \frac{dx}{dt} & = f(t,x), & 
            x(0) & = x_\circ ,
\label{eq:stability_0010}
\end{align} 
the point $x_e$ such that $f(t,x_e) = 0$ is the stationary point of the system described by  (\ref{eq:stability_0010}).
More generally, it is an stationary trajectory as $x(t) = x_e$.
A system is said to be globally asymptotically stable if $x(t) \to x_e$ as $t \to \infty$ for every trajectory $x(t)$.
On the other hand, a system is locally asymptotically stable near or at $x_e$,
if there is an $L > 0$ such that $\bigl|x(0) - x_e\bigr| \leq L \Longrightarrow x(t) \to x_e$ as $t \to \infty$.

Note that in the stability analysis, the linear part of the equation plays an essential role, 
and therefore main investigation is focused on $A(t)  = \frac{\partial{f(x,t)}}{\partial{x}}\Bigr|_{x=x_e} $.  
For the time invariant case the condition $Re{A} < 0$ implies the system is asymptotically stable at $x_e$.

In the case of stochastic differential equations, one has to consider stochastic stability. 
Again, there are various versions of stochastic stability, which include stable in probability, in p-moment, etc.
 \cite{Khasminskii2012}.
Our main interest is the p-moment stability of the Langevin equation with random damping as given by
(\ref{eq:OUMN_0010}). 
In the deterministic case, it is clear that for $\mu(t) > 0$ the system is asymptotically stable at 
$x_e(t) = \eta(t)/\mu(t)$ by Lyapunov condition.
For the stochastic process given by (\ref{eq:OUMN_0010}),
both $\eta$ and $\mu$ are random variables and the problem becomes more complex.

Since the stationary trajectory by itself fluctuates around the mean $\overline{x}_e(t) = 0$, 
in general, it is thus not expected to have p-moment
$[x(t) - x_e(t)]^p$ to approach zero as  $t \to \infty$.   
In fact, 
since the covariance of the random variable $\eta$ is given by a generalized function 
(delta function), which makes the problem even more complicated.

Recall that the random variable $\eta$ is physically linked to thermal fluctuation, and it is temperature dependent.
In fact the covariance of $\eta$ have linear term in temperature $k_BT$ as coefficient, 
which has not been included explicitly in our earlier discussion. 
Therefore at zero temperature, the random variable $\eta$ vanishes on the right hand side of the equation 
(\ref{eq:OUMN_0010}). 
In this case, we have zero temperature time independent stationary point as $x_e = 0$.
However, in case $\mu(t) = 0$, one gets the constant solution $x_e = x_\circ$, which is not of interest to us.

Next, we want to discuss the effect of multiplicative noise on the stability of the Ornstein-Uhlenbeck process. 
A process is said to be stable in the mean and variance if its mean and variance are respectively converge to zero as time goes to infinity. 
From the results in section \mbox{\ref{sec:OUDin}} we conclude that the Ornstein-Uhlenbeck process with damping term given by multiplicative dichotomous noise is not stable since the mean and variance of the resulting process diverge in the long-time limit. 
We remark that instability in the mean and variance also implies energetic instability or instability in the mean energy
\mbox{\cite{MendezHorsthemkeMestresEtAl2011}}

However, if the dichotomous damping noise is introduced as a perturbed term, 
we have shown that both the mean and the variance are bounded for very large time in section \mbox{\ref{sec:OUDin}}.
Thus the process is stable in the mean and variance, hence energetically stable when the random  damping is expressed in terms of original unperturbed constant damping plus the dichotomous noise as the perturbation.

Similarly, it is quite clear from the results in section \mbox{\ref{sec:OUfGN}} that Ornstein-Uhlenbeck process with fractional Gaussian noise as damping is unstable both in mean and covariance. 
Again, it can be stabilized by similar procedure as in the case of dichotomous noise, 
that is by introducing fractional Gaussian noise as a perturbation to the original constant damping. 
We have shown in the last section that both the mean and variance both decay at large time for all 
$0 < H < 1/2$, so the associated process is stable in mean and  bounded in variance.
On the
other hand, for $1/2 < H < 1$, both the mean and variance diverge exponentially as roughly as
$e^{-\mu_\circ{t} + \epsilon^2|t|^{2H}} \sim e^{\epsilon^2|t|^{2H}} $.
At the critical value $H = 1/2$ is they may diverge or converge depending on
$\mu_\circ$ and $\epsilon^2$.

We shall next proceed to calculate explicitly the Lyapunov exponent which is given by
\begin{align}
  \theta_p & = \lim_{t \to \infty} \frac{1}{t}\log\bigl<x(t)^p\bigr>.
\label{eq:stability_0020}
\end{align}
First we note that the solution of 
(\ref{eq:OUMN_0010})
or the Ornstein-Uhlenbeck process with multiplicative noise as damping, can be decomposed into two part as follows:
\begin{align}
  x(t) & = x_\bullet + x_1(t).
\end{align}
Here, the $x_\bullet(t) = x_\circ G(t)$ is non-thermal, 
i.e.\ independent of thermal noise $\eta$, and instead it depends only on the multiplicative noise $\mu(t)$.
On the other hand, $x_1(t) = \int_0^t G(t|u)\eta(u)du$ is the thermal part, it depends on both noises.   
Denote the p-moment by
\begin{align}
  M_p & = M_p^\bullet + M_p^{\text{thermal}} ,
\end{align}
where $M_p^\bullet$ is the p-moment of $x_\bullet(t)$, or the  zero temperature p-moment, 
and $M_p^{\text{thermal}}$ is the contribution from $x_1(t)$, or the non-zero temperature contribution of p-moment. 
The thermal part $M_p^{\text{thermal}}$ involves some power of $x_\bullet$ and $x_1$, in the form
$\Bigl<\bigl[x_{\bullet}(t)\bigr]^{p_\bullet}\bigl[x_1(t)\bigr]^{p_1}\Bigr>$
where $p_1 \geq 1$ and $p_\bullet + p_1 = p$.

We would like to consider first with the zero temperature p-moment, which is simply of the
form
\begin{align}
  M_p^\bullet & = x_\circ^p\bigl<G^p(t)\bigr>.
\label{eq:Tzero_0020}
\end{align}
Based on the results obtained in previous section, we know that all cases under consideration have asymptotic exponential form, either positive or negative.

Here we consider only the processes where the multiplicative noise is introduced as the perturbative term. 
For the dichotomous case it is quite obvious that its first moment is stable.
Unfortunately, the higher moments need more elaboration to ensure their stability. 
However, this is not the case for the fractional Gaussian noise as it takes values from negative infinity to positive infinity. 
We find that it is also asymptotically stable.
For dichotomous noise, it is immediately clear that the p-moment is given by
\begin{align}
  M_{p+}^\bullet & = x_\circ^p e^{-p\mu_\circ{t}}\overline{U^p}(t),    
\label{eq:Tzero_0030}
\end{align}
where $\overline{U^p}$ is similar to $\overline{U}$, except replace $\epsilon$ by $p\epsilon$.
Thus we have for $p\mu_\circ > \alpha_p$, that the Lyapunov exponent is
\begin{align}
  \theta_p & = -(p\mu_\circ - \alpha_p).
\label{eq:Tzero_0040}
\end{align}
It is not obvious that $\theta_p$ is negative for all $p=1,2,\cdots$.
For $p > 0$ and $\mu_\circ^2 - \epsilon^2 > 0$, we know that 
\begin{align}
  (p\mu_\circ + \lambda)^2 - (\lambda^2 + p^2\epsilon^2) & = p^2(\mu_\circ^2 - \epsilon^2) + 2p\lambda\mu > 0.
\label{eq:Tzero_0060}
\end{align}
Thus, we obtain
\begin{align}
  p\mu_\circ - \alpha_p & = p\mu_\circ + \lambda - \sqrt{\lambda^2 + p^2\epsilon^2} > 0 .
\label{eq:Tzero_0050}
\end{align}
It is thus verified that p-moment is stable with the corresponding Lyapunov exponent given by
(\ref{eq:Tzero_0040}).

For the fractional Gaussian noise, the p-moment is given by
\begin{align}
  M_{pg+}^\bullet & = x_\circ^p e^{-p\mu_\circ + p^2\epsilon^2|t|^{2H}}.
\label{eq:Tzero_0070}
\end{align}
Thus it is stable for all $\epsilon$ provided $H < 1/2$, and its Lyapunov exponent is given by
\begin{align}
  \theta_p & = - p\mu_\circ .
\label{eq:Tzero_0080}
\end{align}

For $H=1/2$,  $\xi$ is just the simple Brownian motion, and
\begin{align}
  \theta_p & = -(p\mu_\circ - p^2\epsilon^2) .
\label{eq:Tzero_0090}
\end{align}
This indicates that it is stable only for low value of $p$, for example, $p=1$.

The thermal part of the p-moment involves only even power of $x_1(t)$.
Let us consider the 2-moment. 
\begin{align}
  M_2 & = M_2^\bullet + M_2^1
\end{align}
It is clear that the thermal part is only pure, 
i.e.\ involves only $x_1$:
As we have discussed above the zero-temperature p-moment
$M_p^\bullet \xrightarrow[t\to\infty]{} 0$
for the correspondence conditions. 
From the estimates in previous sections, for either dichotomous and fractional Gaussian noise, 
we have provided the calculation and graph for $K_{1+}(t,s)$, which converges to a constant as $t \to \infty$.
It is clear that the corresponding thermal 2-moment $M_2^{\text{thermal}}(t) = S_{1+}(t)$ converges to a constant. 
Since
\begin{align}
 \theta_p & = \lim_{t\to\infty}\frac{1}{t}\log\bigl[M_p^\bullet(t) + M_p^{\text{thermal}}(t)\bigr]
            = \lim_{t\to\infty}\frac{1}{t}\log\bigl[M_p^{\text{thermal}}(t)\bigr],
\end{align}
which gives
\begin{align}
  \theta_2 & = \theta_2^{\text{thermal}} = 0.
\end{align}
By definition, this implies the process is not Lyapunov asymptotically stable in 2-moment. 
It is obvious that the process is not Lyapunov asymptotically for stable all higher p-moments as well.

For non-zero positive temperature, there exists thermal fluctuation. 
However, the processes are still stable in the mean. 
The fluctuation around the mean is the usual one.
Thus in this case only the 1-moment are consider. 
The Lyapunov exponent of dichotomic noise and fractional Gaussian noise are given
by the same exponent as above for the zero temperature case 1-moment.


\section{Conclusion}
\label{sec:conclusion}
We are able to obtain the explicit solutions of the Langevin equation with random damping given by dichotomous and fractional Gaussian noise. 
The mean and covariance of the resulting processes are calculated explicitly. 
Asymptotic properties of the mean and covariance are studied.
The resulting Ornstein-Uhlenbeck process is unstable in the mean, variance as well as energetic stable if the damping term is given purely by the dichotomous noise. 
However, if the dichotomous noise appears as a perturbation term to the original constant damping, the process becomes stable in the mean, variance and is energetic stable. 
In the case of damping given by fractional Gaussian noise, 
a similar situation exists. 
Direct use of fractional Gaussian noise as the random damping term leads to a Gaussian process which is unstable in mean and variance, hence energetically unstable. 
Again, if the fractional Gaussian noise enters as a perturbative term to the original constant damping, the resulting process is stable in mean, variance and energy. 
On the other hand, the Ornstein-Uhlenbeck processes with random damping considered in this paper 
are Lyapunov asymptotically unstable in 2- and higher moment.
We remark that our results on the fractional Gaussian damping have the potential 
for modeling physical systems experiencing random damping 
since it has the advantage that the Hurst index associated with the fractional Gaussian noise can serve as 
an adjustable parameter for the model.

One possible extension of our study is to allow the mass to fluctuate in addition to the random damping. 
The stochastic equation describing such a system involves two multiplicative noise terms, 
which will be be difficult to solve and one expects analytic solution does not exist.
Burov and Gitterman 
\mbox{\cite{BurovGitterman2016}}
have recently considered such a system with both the multiplicative noises of dichotomous type. 
They used the Shapiro-Loginov formula to obtain the mean and variance and study the stability of the system in the mean and variance.

Another generalization is to consider the fractional version of Langevin equation with random damping term. 
For constant damping, fractional Langevin equation with single and two indices
\mbox{\cite{LimLiTeo2008,LimTeo2009}},
with distributed order \mbox{\cite{EabLim2011}},
and generalized fractional Langevin equation 
\mbox{\cite{LimTeo2009a,EabLim2010}}
have been previously considered.
One would expect the solutions for these fractional generalizations to be even more complex since the simpler cases discussed in this paper show the solutions are rather involved. 
Numerical and approximation methods have to be employed to study such systems.

Finally, one may wish to consider more realistic models by taking into consideration both the fluctuating medium or environment and the effects of time delays 
\cite{NieMei2008}, 
which are of particular interest in population and biomedical dynamics,
economics and finance, chemical and biochemical processes, etc. 
On the other hand, in modeling many complex systems the
effect of aging becomes significant (see for example, 
\cite{SafdariChechkinJafariEtAl2015}). 
The aging effect, a characteristic of non-stationary process, is the
dependence of statistical properties of the process on the time lag between the initiation of the system and the start time of
the observation of the process. A recent study 
\cite{CherstvyVinodAghionEtAl2017}  
has considered the aging and delay time analysis of financial time series.
It would be interesting to carry out a similar study for our system.



\appendix
\renewcommand{\theequation}{A\arabic{equation}}

\section*{Appendix: Calculation of $\Bigl<G(t|u)G(s|u)\Bigr>$}
\label{sec:CalDiNGG}
In this appendix, we give some details on the derivation of 
(\ref{eq:OUDiN_0110}).

Note that
\begin{align}
  g_n(t|s) & = \int_s^t dt_n \cdots \int_s^{t_2}d_1 (-1)^{\sum_{k=1}^n N(t_k)} \nonumber \\
           & = \int_s^t dt_n \cdots \int_s^{t_2}d_1 (-1)^{\sum_{k=1}^n [N(t_k)-N(s)] +nN(s)} \nonumber \\
           & = \int_s^t dt_n \cdots \int_s^{t_2}d_1 (-1)^{\sum_{k=1}^n N(t_k-s) +nN(s)} .
\label{eq:CalDiNGG_0010}
\end{align}
By changing variable $t_k^\prime = t_k -s$ for $k=1,2,\cdots,n$, (\ref{eq:CalDiNGG_0010}) can be rewritten as
\begin{align}
  g_n(t|s) & = \int_0^{t-s} dt_n \cdots \int_0^{t_2-s}d_1 (-1)^{\sum_{k=1}^n N(t_k) +nN(s)} 
             = g_n(t-s)(-1)^{nN(s)} .
\label{eq:CalDiNGG_0020}
\end{align}
Therefore,
\begin{align}
  G(t|s) & = U(t-s) - V(t-s)(-1)^{nN(s)}.
\label{eq:CalDiNGG_0030}
\end{align}

Now, (\ref{eq:OUDiN_0110}) can be expressed as
\begin{align}
  \bigl<G(t|s)G^2(s|u)\bigr> & =  \bigl<U(t-s)\bigr>\bigl<G^2(s|u)\bigr> - \bigl<V(t-s)\bigr>\bigl<(-1)^{N(s)}G^2(s|u)\bigr> .
\label{eq:CalDiNGG_0040}
\end{align}

Furthermore, one notes that
\begin{align}
  (-1)^{N(t)}g_n(t|s) & = \frac{d}{dt}g_{n+1}(t|s),
\label{eq:CalDiNGG_0050}
\end{align}
which can be easily verify by expressing the integral explicitly, then differentiate with respect to the upper limit $t$.
From expression (\ref{eq:OUDiN_0040}) and (\ref{eq:CalDiNGG_0050}), one can write
\begin{align}
    (-1)^{N(s)}G^2(s|u) & = \frac{1}{(-2\epsilon)}\frac{d}{dt}G^2(s|u).
\label{eq:CalDiNGG_0060}
\end{align}
Note that $g_\circ(s|u) = 1$.
Inserting (\ref{eq:CalDiNGG_0060}) in (\ref{eq:CalDiNGG_0040}) we obtain (\ref{eq:OUDiN_0110}).




\bibliographystyle{elsarticle-num}

\bibliography{biblioOUFDxx}

\begin{thebibliography}{10}
\expandafter\ifx\csname url\endcsname\relax
  \def\url#1{\texttt{#1}}\fi
\expandafter\ifx\csname urlprefix\endcsname\relax\def\urlprefix{URL }\fi
\expandafter\ifx\csname href\endcsname\relax
  \def\href#1#2{#2} \def\path#1{#1}\fi

\bibitem{Dyson53}
F.~J. Dyson, The dynamics of a disordered linear chain, Phys. Rev. 92 (1953)
  1331--1338.

\bibitem{Leibowitz63}
M.~A. Leibowitz, Statistical behavior of linear systems with randomly varying
  parameters, J. Math. Phys. 4~(6) (1963) 852--858.

\bibitem{Frisch68}
U.~Frisch, Probabilistic Methods in Applied Mathematics, Vol.~I, Academic
  Press, New York, 1968, p.~75.

\bibitem{BourretFrischPouquet73}
R.~C. Bourret, U.~Frisch, A.~Pouquet, Brownian motion of harmonic oscillator
  with stochastic frequency, Physica 65~(2) (1973) 303--320.

\bibitem{KlyatskinTatarskii74}
V.~I. Klyatskin, V.~I. Tatarski\v{\i}, Diffusive random process approximation
  in certain nonstationary statistical problems of physics, Phys. Usp. 16~(4)
  (1974) 494--511.

\bibitem{VanKampen76}
N.~G. {V}an {K}ampen, Stochastic differential equations, Phys. Rep. 24~(3)
  (1976) 171 -- 228.

\bibitem{Gitterman14}
M.~Gitterman, Oscillator and Pendulum with a Random Mass, World Scientific
  Singapore, 2015.

\bibitem{Luczka95}
J.~{\L}uczka, P.~H{\"a}nggi, A.~Gadomski, Diffusion of clusters with randomly
  growing masses, Phys. Rev. E 51 (1995) 5762--5769.

\bibitem{AusloosLambiotte2006a}
M.~Ausloos, R.~Lambiotte, Brownian particle having a fluctuating mass, Phys.
  Rev. E 73 (2006) 011105.

\bibitem{BlumWurmKempfEtAl2000}
{J. Blum et al.}, Growth and form of planetary seedlings: Results from a
  microgravity aggregation experiment, Phys. Rev. Lett. 85 (2000) 2426--2429.

\bibitem{BlumWurm00}
J.~Blum, G.~Wurm, Experiments on sticking, restructuring, and fragmentation of
  preplanetary dust aggregates, Icarus 143~(1) (2000) 138--146.

\bibitem{WeidenschillingEtAl97}
S.~J. Weidenschilling, D.~Spaute, D.~R. Davis, F.~Marzari, K.~Ohtsuki,
  Accretional evolution of a planetesimal swarm, Icarus 128~(2) (1997) 429 --
  455.

\bibitem{KuipersBarkema2009}
J.~Kuipers, G.~T. Barkema, Non-markovian dynamics of clusters during
  nucleation, Phys. Rev. E 79 (2009) 062101.

\bibitem{Kaiser02}
N.~Kaiser, Review of the fundamentals of thin-film growth, Appl. Opt. 41~(16)
  (2002) 3053--3060.

\bibitem{PerezSavilleSoria01}
A.~T. P{\'r}ez, D.~Saville, C.~Soria, Modeling the electrophoretic deposition
  of colloidal particles, Europhys. Lett. 55~(3) (2001) 425.

\bibitem{Nagatani96}
T.~Nagatani, Kinetics of clustering and acceleration in 1d traffic flow, J.
  Phys. Soc. Jpn. 65~(10) (1996) 3386--3389.

\bibitem{Ben-NaimKrapivskyRedner1994}
E.~Ben-Naim, P.~L. Krapivsky, S.~Redner, Kinetics of clustering in traffic
  flows, Phys. Rev. E 50 (1994) 822--829.

\bibitem{Ausloos2002}
M.~Ausloos, K.~Ivanova, Mechanistic approach to generalized technical analysis
  of share prices and stock market indices, Eur. Phys.J. B 27~(2) (2002)
  177--187.

\bibitem{Canessa2009}
E.~Canessa, Stock market and motion of a variable mass spring, Physica A
  388~(11) (2009) 2168--2172.

\bibitem{Klyatskin2005}
V.~I. Klyatskin, Dynamics of Stochastic Systems, Elsevier, Amsterdam, 2005.

\bibitem{Tatarski61}
V.~I. Tatarski\v{\i}, {Wave Propagation in Turbulent Medium}, McGraw-Hill,
  1961.

\bibitem{Phillips1977}
O.~M. Phillips, The Dynamics of the Upper Ocean, 2nd Edition, Cambridge
  University, 1977.

\bibitem{WestSeshadri1981}
B.~J. West, V.~Seshadri, Model of gravity wave growth due to fluctuations in
  the air-sea coupling parameter, J. Geophys. Res. Oceans 86~(C5) (1981)
  4293--4298.

\bibitem{TakayasuSatoTakayasu1997}
H.~Takayasu, A.-H. Sato, M.~Takayasu, Stable infinite variance fluctuations in
  randomly amplified {L}angevin systems, Phys. Rev. Lett. 79 (1997) 966--969.

\bibitem{Turelli1977}
M.~Turelli, Random environments and stochastic calculus, Theor. Popul. Biol.
  12~(2) (1977) 140 --178.

\bibitem{VitanovarXiv2013}
N.~K. {Vitanov}, K.~N. {Vitanov}, {On systems of interacting populations
  influenced by multiplicative white noise}, aeXiv:1311.3567 [nlin.CD] (2013).

\bibitem{VitanovVitanov2014}
N.~K. Vitanov, K.~N. Vitanov, Population dynamics in presence of state
  dependent fluctuations, Comput. Math. Appl. 68~(9) (2014) 962--971.

\bibitem{DykmanKhasinPortmanEtAl2010}
M.~I. Dykman, M.~Khasin, J.~Portman, S.~W. Shaw, Spectrum of an oscillator with
  jumping frequency and the interference of partial susceptibilities, Phys.
  Rev. Lett. 105 (2010) 230601.

\bibitem{ZhangMoserGuettingerEtAl2014}
Y.~Zhang, J.~Moser, J.~G\"uttinger, A.~Bachtold, M.~I. Dykman, Interplay of
  driving and frequency noise in the spectra of vibrational systems, Phys. Rev.
  Lett. 113 (2014) 255502.

\bibitem{MiaoYeomWangEtAl2014}
T.~Miao, S.~Yeom, P.~Wang, B.~Standley, M.~Bockrath, Graphene
  nanoelectromechanical systems as stochastic-frequency oscillators, Nano Lett.
  14~(6) (2014) 2982--2987.

\bibitem{SnyderJoshiSerna2014}
P.~Snyder, A.~Joshi, J.~D. Serna, Modeling a nanocantilever-based biosensor
  using a stochastically perturbed harmonic oscillator, Int. J. Nanosci.
  13~(02) (2014) 1450011.

\bibitem{SunZouMaizelisEtAl2015}
F.~Sun, J.~Zou, Z.~A. Maizelis, H.~B. Chan, Telegraph frequency noise in
  electromechanical resonators, Phys. Rev. B 91 (2015) 174102.

\bibitem{SansaSageBullardEtAl2016}
{M. Sansa et al.}, Frequency fluctuations in silicon nanoresonators, Nat. Nano
  11~(6) (2016) 552--558.

\bibitem{Simon79}
B.~Simon, Functional Integration and Quantum Physics, Academic Pres , New York,
  1979, p.~35.

\bibitem{Roepstorff94}
G.~Roepstorff, Path Integral Approach to Quantum Physics: An Introduction,
  Springer, Berlin, 1994, p.~36.

\bibitem{Dimock11}
J.~Dimock, Field Theory: A Mathematical Primer, Cambridge University Press,
  Cambridge, 2011, p. 169.

\bibitem{Casteren15}
{Jan A. {V}an~{C}asteren}, Advanced Stochastic Processes: Part I, second
  edition Edition, Bookboon, 2015, p.~98.

\bibitem{DantasPedrosaBaseia1992}
C.~M.~A. Dantas, I.~A. Pedrosa, B.~Baseia, Harmonic oscillator with
  time-dependent mass and frequency and a perturbative potential, Phys. Rev. A
  45 (1992) 1320--1324.

\bibitem{DantasPedrosaBaseia1992a}
C.~M.~A. Dantas, I.~A. Pedrosa, B.~Baseia, Harmonic oscillator with
  time-dependent mass and frequency, Braz. J. Phys. 22~(1) (1992) 33--39.

\bibitem{KanasugiOkada1995}
H.~Kanasugi, H.~Okada, Systematic treatment of general time-dependent harmonic
  oscillator in classical and quantum mechanics, Progr. Theor. Phys. 93~(5)
  (1995) 949--960.

\bibitem{ChaJangJungEtAl2015}
J.~Cha, et~al., On the exact solutions of the damped harmonic oscillator with a
  time-dependent damping constant and a time-dependent angular frequency, J.
  Korean Phys. Soc. 67~(2) (2015) 404--408.

\bibitem{JasnowGerjuoy1975}
D.~Jasnow, E.~Gerjuoy, Langevin equation with stochastic damping. {P}ossible
  application to critical binary fluid, Phys. Rev. A 11 (1975) 340--349.

\bibitem{GrueOeksendal1997}
J.~Grue, B.~{\O}ksendal, A stochastic oscillator with time-dependent damping,
  Stoch. Proc. Appl. 68~(1) (1997) 113 -- 131.

\bibitem{LuczkaTalknerHaenggi2000}
J.~{\L}uczka, P.~Talkner, P.~H{\"a}ngg, Diffusion of {B}rownian particles
  governed by fluctuating friction, Physica A 278~(1–2) (2000) 18 -- 31.

\bibitem{Gitterman2004}
M.~Gitterman, Harmonic oscillator with fluctuating damping parameter, Phys.
  Rev. E 69 (2004) 041101.

\bibitem{Gitterman2005}
M.~Gitterman, Classical harmonic oscillator with multiplicative noise, Physica
  A 352 (2005) 309 -- 334.

\bibitem{LeprovostAumaitreMallick2006}
N.~Leprovost, S.~Auma{\^{\i}}tre, K.~Mallick, Stability of a nonlinear
  oscillator with random damping, Eur. Phys. J. B 49~(4) (2006) 453--458.

\bibitem{MendezHorsthemkeMestresEtAl2011}
V.~M\'endez, W.~Horsthemke, P.~Mestres, D.~Campos, Instabilities of the
  harmonic oscillator with fluctuating damping, Phys. Rev. E 84 (2011) 041137.

\bibitem{WestLindenbergSeshadri1980}
B.~J. West, K.~Lindenberg, V.~Seshadri, Brownian motion of harmonic systems
  with fluctuating parameters. {I.} {E}xact first and second order statistics
  of a mechanical oscillator, Physica A 102~(3) (1980) 470 -- 488.

\bibitem{Lindenberg1980}
K.~Lindenberg, V.~Seshadri, K.~E. Shuler, B.~J. West, Equal-time second-order
  moments of a harmonic oscillator with stochastic frequency and driving force,
  J. Stat. Phys. 23~(6) (1980) 755--765.

\bibitem{LindenbergSeshadriWest1980}
K.~Lindenberg, V.~Seshadri, B.~J. West, Brownian motion of harmonic systems
  with fluctuating parameters. {II.} {R}elation between moment instabilities
  and parametric resonance, Phys. Rev. A 22 (1980) 2171--2179.

\bibitem{LindenbergSeshadriWest1981}
K.~Lindenberg, V.~Seshadri, B.~J. West, Brownian motion of harmonic systems
  with fluctuating parameters. {III.} {S}caling and moment instabilities,
  Physica A 105~(3) (1981) 445 -- 471.

\bibitem{Gitterman2003}
M.~Gitterman, Harmonic oscillator with multiplicative noise: Nonmonotonic
  dependence on the strength and the rate of dichotomous noise, Phys. Rev. E 67
  (2003) 057103.

\bibitem{MankinLaasLaasEtAl2008}
R.~Mankin, K.~Laas, T.~Laas, E.~Reiter, Stochastic multiresonance and
  correlation-time-controlled stability for a harmonic oscillator with
  fluctuating frequency, Phys. Rev. E 78 (2008) 031120.

\bibitem{ShapiroLoginov1978}
V.~E. Shapiro, V.~M. Loginov, Formulae of differentiation and their use for
  solving stochastic equations, Physica A 91~(3) (1978) 563 -- 574.

\bibitem{Kampen1992}
N.~G. {V}an {K}ampen, Stochastic Processes in Physics and Chemistry, Elsevier,
  Amsterdam, 1992.

\bibitem{MandelbrotNess1968}
B.~B. Mandelbrot, J.~W. {Van Ness}, Fractional {B}rownian motions, fractional
  noises and applications, SIAM Rev. 10~(4) (1968) 422--437.

\bibitem{Samorodnitsky1994}
G.~Samorodnitsky, M.~S. Taqqu, Stable Non-Gaussian Random Processes: Stochastic
  Models with Infinite Variance, Chapman \& Hall, New York, 1994.

\bibitem{GradshteynRyzhik1980}
I.~S. Gradshteyn, I.~M. Ryzhik, Table of Integrals, Series, and Products,
  Academic Press, New York, 1980.

\bibitem{AnishchenkoVadivasovaStrelkova2014}
V.~S. Anishchenko, T.~E. Vadivasova, G.~I. Strelkova, Deterministic Nonlinear
  Systems, Springer International Publishing Switzerland, 2014.

\bibitem{Khasminskii2012}
R.~Khasminskii, Stochastic Stability of Differential Equations, Springer New
  York, 2012.

\bibitem{BurovGitterman2016}
S.~Burov, M.~Gitterman, The noisy oscillator : Random mass and random damping,
  arXiv:1607.06289v2 [cond-mat.stat-mech].

\bibitem{LimLiTeo2008}
S.~C. Lim, M.~Li, L.~P. Teo, Langevin equation with two fractional orders,
  Phys. Lett. A 372~(42) (2008) 6309 -- 6320.

\bibitem{LimTeo2009}
S.~C. Lim, L.~P. Teo, The fractional oscillator process with two indices, J.
  Phys. A: Math. Theor. 42~(6) (2009) 065208.

\bibitem{EabLim2011}
C.~H. Eab, S.~C. Lim, Fractional {L}angevin equations of distributed order,
  Phys. Rev. E 83 (2011) 031136.

\bibitem{LimTeo2009a}
S.~C. Lim, L.~P. Teo, Modeling single-file diffusion with step fractional
  {B}ownian motion and a generalized fractional {L}angevin equation, J. Stat.
  Mech. Theor. Exp. 2009~(08) (2009) P08015.

\bibitem{EabLim2010}
C.~H. Eab, S.~C. Lim, Fractional generalized {L}angevin equation approach to
  single-file diffusion, Physica A 389~(13) (2010) 2510 -- 2521.

\bibitem{NieMei2008}
L.~Nie, D.~Mei, Effects of time delay on symmetric two-species competition
  subject to noise, Phys. Rev. E 77 (2008) 031107.

\bibitem{SafdariChechkinJafariEtAl2015}
H.~Safdari, A.~V. Chechkin, G.~R. Jafari, R.~Metzler, Aging scaled brownian
  motion, Phys. Rev. E 91 (2015) 042107.

\bibitem{CherstvyVinodAghionEtAl2017}
A.~Cherstvy, D.~Vinod, E.~Aghion, A.~V. Chechkin, R.~Metzler, Time averaging,
  ageing and delay analysis of financial time series, New J. Phys. 19 (2017)
  063045.

\end{thebibliography}

\end{document}